\documentclass[conference]{IEEEtran}
\IEEEoverridecommandlockouts
\usepackage{cite}
\usepackage{amsmath,amssymb,amsfonts}
\usepackage{algorithmic}
\usepackage{graphicx}
\usepackage{textcomp}
\usepackage{xcolor}
\usepackage{tabularx}
\usepackage{hyperref}
\def\BibTeX{{\rm B\kern-.05em{\sc i\kern-.025em b}\kern-.08em
    T\kern-.1667em\lower.7ex\hbox{E}\kern-.125emX}}
\begin{document}

\title{BC-MRI-SEG: A Breast Cancer MRI Tumor Segmentation Benchmark \\
\thanks{}
}

\author{Anthony Bilic, Chen Chen \\
Center for Research in Computer Vision, University of Central Florida \\
\tt\small 
an609701@ucf.edu; chen.chen@crcv.ucf.edu}

\maketitle

\begin{abstract}
Binary breast cancer tumor segmentation with Magnetic Resonance Imaging (MRI) data is typically trained and evaluated on private medical data, which makes comparing deep learning approaches difficult. We propose a benchmark (BC-MRI-SEG) for binary breast cancer tumor segmentation based on publicly available MRI datasets. The benchmark consists of four datasets in total, where two datasets are used for supervised training and evaluation, and two are used for zero-shot evaluation. Additionally we compare state-of-the-art (SOTA) approaches on our benchmark and provide an exhaustive list of available public breast cancer MRI datasets. The source code has been made available at \url{https://irulenot.github.io/BC_MRI_SEG_Benchmark/}.

\end{abstract}


\section{Introduction}
Tumor Segmentation is the task of locating a tumor in an image and labeling each pixel as tumor or background. The Brain Tumor Segmentation (BraTS) \cite{b3} benchmark is popular for this task and contains 251 patients. For breast cancer, The Reference Image Database to Evaluate Therapy Response Breast MRI (RIDER) \cite{b28} benchmark is most commonly used but is limited due to only having five patients. The lack of labeled data is prevalent in medical imaging \cite{b21}, and there is a need for more extensive and more diverse datasets \cite{b21}. Additionally, the field requires robust and distribution adaptable models for clinical settings \cite{b27}.

To address these needs, we created the Breast Cancer MRI Segmentation Benchmark (BC-MRI-SEG), consisting of four public datasets, including RIDER \cite{b28}, totaling 1,320 patients. Two datasets are used for training, and two are used for zero-shot evaluation.

The segmentation datasets used for training include ISPY1-Tumor-SEG-Radiomics (ISPY1) \cite{b8} and BreastDM \cite{b45}. The datasets used for zero-shot evaluation include The Reference Image Database to Evaluate Therapy Response Breast MRI (RIDER) \cite{b28} and Dynamic contrast-enhanced magnetic resonance images of breast cancer patients with tumor locations (DUKE) \cite{b34}.

Four architectures are compared. The models evaluated include Convolutional Networks for Biomedical Image Segmentation (U-Net2D) \cite{b33}, Three dimensional (3D) MRI brain tumor segmentation using autoencoder regularization (SegResNet) \cite{b30}, 3. Swin Transformers for Semantic Segmentation of Brain Tumors in MRI Images (SwinUNETR) \cite{b17}, and Adapting Segment Anything Model for Medical Image Segmentation (Med-SA) \cite{b42}

\indent Our contributions include the following:
\begin{enumerate}
    \item We collect, prepare, and share four public breast cancer MRI datasets that the medical imaging community can easily access.
    \item We present the most comprehensive list of available public breast cancer MRI data, including nine distinct datasets with multiple modalities.
    \item We present the first unified breast cancer MRI segmentation benchmark, which evaluates a model's ability to generalize across different breast cancer MRI datasets. We call this benchmark BC-MRI-SEG.
    \item We provide a comparison of SOTA deep learning approaches, specifically in the domain of breast cancer MRI tumor segmentation.
\end{enumerate}

\section{Related Works}

Few works specifically focus on breast cancer tumor segmentation using MRI data \cite{b5, b6, b14, b16, b32}, and most train and evaluate their approaches on internal datasets or have fewer than 50 patients in their training sets. BTS-GAN \cite{b16} trains and evaluates a conditional GAN approach on the RIDER \cite{b28} dataset. Park et al. \cite{b32} apply the UNETR model \cite{b46} on an internal dataset with 736 patients. Bouchebbah et al. \cite{b5} propose a two-stage automatic tumor segmentation method on a private dataset with 18 patients and also on the RIDER \cite{b28} dataset. Yin et al. \cite{b14} combine the level set and ray casting algorithms for a three-step approach on a private dataset with ten patients.

These works focus on proposing methods that achieve reasonable results on their respective datasets but do not test their approaches' zero-shot capabilities. In addition, it is difficult to compare their effectiveness due to their code and datasets not being publicly available.

BreastDM \cite{b45} releases a high-quality breast cancer MRI dataset with 232 patients and evaluates the performance of various models on the classification and segmentation tasks. They find that U-Net performs the best on their segmentation task, and we integrate their dataset with our benchmark.  

\section{Datasets}
The datasets used (ISPY1 \cite{b8}, BreasDM \cite{b45}, RIDER \cite{b28}, and DUKE \cite{b34}) in this benchmark, whose details are shown in Table \ref{tab:1}, are aggregated from four different medical studies. Each study varied in procurement method and primarily differed in what MRI scanners are used, their scanner configurations, and how the imaging data is processed after collection. ISPY1 \cite{b8}, BreastDM \cite{b45}, and RIDER \cite{b28} contain expert-labeled binary 3D tumor segmentations, while DUKE \cite{b34} has expert-labeled 3D box annotations around the tumors. The differences in procurement pose the challenge for deep learning approaches to be both robust and generalizable.

Each dataset's images differ in content and the number of channels they contain. ISPY1 \cite{b8} consists of three channels, and their image acquisition protocol includes a localization scan, a T2-weighted sequence, and a contrast-enhanced T1-weighted series. BreastDM \cite{b45} also consists of three channels: a pre-contrast sequence, a post-contrast sequence, and the subtracted sequences (obtained by subtracting the second post-contrast from the pre-contrast sequences). RIDER \cite{b28} consists of four channels: an early anatomical reference, ADC, B0, and B800. Finally, a single fat-saturated gradient echo T1-weighted pre-contrast channel is used for DUKE \cite{b34}.

Public breast cancer MRI data generally comes from studies that monitor the tumors before and after treatments and are paired with clinical and outcome data. In this benchmark only scans taken before treatment are used. For training and evaluation, the sequences are reshaped to be in \((D, C, H, W)\) format, where \(D\) is the depth, \(C\) is the number of channels, and \(H\) and \(W\) are the height and width of the images. Scripts for preparing each dataset are provided in our \href{https://irulenot.github.io/BC_MRI_SEG_Benchmark/}{GitHub}.

Outside of the datasets used in this benchmark, we present an extensive list of public breast cancer MRI datasets in Table \ref{tab:2}, which contain multiple modalities and are complemented by text annotations from \cite{b18}. We note that while the TCGA \cite{b29} and ISPY2 \cite{b25} datasets have segmentation masks, we struggled to utilize them due to inconsistent pairings and poor annotations.

\begin{table}[htbp]
    \centering
    \small
    \setlength{\tabcolsep}{2pt}
    \caption{Benchmark Datasets}
    \begin{tabularx}{\linewidth}{|X|c|c|c|c|c|}
        \hline
        \textbf{Dataset} & \textbf{Labels} & \textbf{Patients} & \textbf{Scans} & \textbf{Images} & \textbf{Annotations} \\
        \hline
        ISPY1 \cite{b8} & SEG & 161 & 483 & 67,080 & 6,801 \\
        BreastDM \cite{b45} & SEG & 232 & 696 & 4,095 & 4,095 \\
        RIDER \cite{b28} & SEG & 5 & 20 & 1,200 & 90 \\
        DUKE \cite{b34} & BOX & 922 & 922 & 157,198 & 23,426 \\
        \hline
        \multicolumn{6}{l}{\parbox{\dimexpr\linewidth-2\tabcolsep}{Quantitative details about the datasets used in the BC-MRI-SEG benchmark. Each dataset contains binary breast cancer tumor labels. SEG denotes segmentation. BOX denotes bounding box.}} \\
    \end{tabularx}
    \label{tab:1}
\end{table}

\begin{table}[htbp]
    \centering
    \small
    \setlength{\tabcolsep}{2pt}
    \caption{Available Breast Cancer MRI Datasets}
    \begin{tabularx}{\linewidth}{|X|c|c|c|}
        \hline
        \textbf{Dataset} & \textbf{Patients} & \textbf{Labels} & \textbf{Modalities} \\
        \hline
        ISPY2 \cite{b25} & 361 & SEG & MR \\
        TCGA \cite{b29} & 139 & SEG & MR, MG, PT \\
        DIAGNOSIS \cite{b4} & 88 & CP & MR, MG, CT, PT \\
        QIN-Breast \cite{b26} & 68 & N/A & MR, CT, PT \\
        NACT-Pilot \cite{b31} & 64 & SEG & MR \\
        \hline
        \multicolumn{4}{l}{\parbox{\dimexpr\linewidth-2\tabcolsep}{Quantitative details about public breast cancer MRI datasets which are not included in the the BC-MRI-SEG benchmark. For the labels SEG denotes segmentation and C/P denotes center points. For the modalities MR denotes MRI, MG denotes mammograph, PT denotes pathology, and CT denotes computed tomography. }} \\
    \end{tabularx}
    \label{tab:2}
\end{table}

\section{Evaluation Method}
DICE (DSC) \cite{b2} and F1 \cite{b19} scores are used to evaluate various models such as in other SOTA works \cite{b31, b41}. DICE score is used in the case of 3D segmentation and can be formulated as, \( DICE(A, B) = \frac{2 \cdot |A \cap B|}{|A| + |B|} \) where \(A\) and \(B\) are the predicted and true segmentation masks. F1 score is used in the case of 3D classification and can be formulated as \( F1(C, D) = \frac{2 \cdot \text{precision}(C, D) \cdot \text{recall}(C, D)}{\text{precision}(C, D) + \text{recall}(C, D)} \) where \(C\) and \(D\) are the slices classified as tumor present and tumor absent.

Benchmark evaluation is done in two stages. The first stage consists of supervised training and evaluation with the ISPY1 \cite{b8} and BreastDM \cite{b45} datasets. The two datasets are aggregated and then used for evaluation, as seen in Table \ref{tab:4}. The second stage consists of performing zero-shot evaluation on the RIDER \cite{b28} and DUKE \cite{b34} datasets as seen in Table \ref{tab:5}. Note that the evaluation task for DUKE \cite{b34} is classification as is done by \cite{b20} and formally defined in the previous paragraph. Zero-shot performance is significant because it demonstrates a model architecture's ability to generalize to new and unseen data. \cite{b15}

\section{Experiment}

The training paradigm used leverages Dice Loss \cite{b2}, the AdamW optimizer \cite{b23}, and a CosineAnnealing scheduler \cite{b23} as is seen in other SOTA training schemes \cite{b31, b41}. Spatial crop, channel flip, channel-wise normalization, scaled intensity, and shift intensity augmentations are performed during training orientation. Both ISPY1 \cite{b8} and BreastDM \cite{b45} are separately split into 80\%-20\% training-test patient splits and then respectively combined. Splitting the data by patients ensures evaluation of the model's ability to generalize across different patients, which is crucial in clinical settings. We train each model with one MRI sequence at a time. 

Benchmark datasets are prepared by clipping the top 0.01\% of pixel values \cite{b45}, performing Z-normalization \cite{b45}, and resizing the sequences to have a depth of 128 and a width and height of 256; such as is done by SOTA methods \cite{b10}.
Because the training datasets, ISPY1 \cite{b8} and BreastDM \cite{b45}, have three channels, the zero-shot datasets are modified.
DUKE \cite{b34}, which only has a single channel denoted as \( X \), is expanded to have three channels. \(\ X_{\text{expanded}} = [X, X, X]\).
For RIDER \cite{b28} only the first channel is used. \(\ X_{\text{RIDER}} = [X_{\text{RIDER}}(1, :, :), X_{\text{RIDER}}(1, :, :), X_{\text{RIDER}}(1, :, :)].\) Only using RIDER's \cite{b28} first channel is based on our empirical results but we encourage the use of all four channels.

We choose the U-Net architecture as our baseline model due to its widespread use in medical imaging \cite{b43}. Three variations are tested: U-Net2D \cite{b33}, U-Net3D \cite{b22}, and U-Net2.1D. U-Net3D \cite{b22} extends the original U-Net2D \cite{b33}, which replaces 2D convolutions and pooling layers with their 3D counterparts. 

Note that 3D segmentation models process an entire MRI sequence ($\mathbb{R}^{D \times H \times W \times C_{\text{in}}}$) and output a mask volume ($\mathbb{V}_{\text{3D}} \in \mathbb{R}^{D' \times H' \times W' \times C_{\text{out}}}$), while 2D models label each image individually ($\mathbb{I}_{\text{2D}} \in \mathbb{R}^{H \times W \times C_{\text{in}}}$). Where $D$, $H$, $W$ are the dimensions of depth, height, and width respectively, and $C$ is the number of channels. 

U-Net2.1D is an approach we propose and is similar to U-Net2D \cite{b33} except that it has nine input channels (\( C_{\text{in}} = 9 \)) instead of three (\( C_{\text{in}} = 3 \)). This increase in channels allows the model to ingest three sets of three-channel images, represented as \( I_{\text{prior}} \), \( I_{\text{current}} \), and \( I_{\text{subsequent}} \), corresponding to the prior, current, and subsequent MRI images. 

\[ U\text{-}Net2.1D: \mathbb{R}^{H \times W \times C_{\text{in}}} \rightarrow \mathbb{R}^{H' \times W' \times C_{\text{out}}} \]

\( H \) and \( W \) denote the height and width of the input images, while \( C_{\text{out}} \) represents the output channels of the network. Specifically, a segmentation mask is only output for the center image (\( I_{\text{current}} \)).

Three SOTA 3D approaches are also evaluated. They include SwinUNETR \cite{b17}, Med-SA \cite{b42}, and SegResNet \cite{b30}. SwinUNETR \cite{b17} utilizes a U-shaped network with a Swin transformer \cite{b24} as the encoder and connects it to a CNN-based decoder at different resolutions via skip connections. Med-SA \cite{b42} incorporates domain-specific medical knowledge into the Segment Anything (SAM) model using an adapter finetuning technique. SegResNet \cite{b30} is an encoder-decoder architecture with an asymmetrically large encoder.

The validity of our training approach is substantiated through training the 3D U-Net model (U-Net3D) \cite{b22} on two popular datasets, BraTS \cite{b3} and MSD Spleen \cite{b46}, in which we replicate the mean dice scores of 0.71 and 0.90 respectively. We perform an ablation with the U-Net3D \cite{b22} model on the benchmark's segmentation datasets ISPY1 \cite{b8}, BreastDM \cite{b45}, and RIDER \cite{b28}. This ablation is done to indirectly evaluate the datasets and later compare the performance of U-Net3D \cite{b22} when it is trained on multiple datasets. Results are shown in Table \ref{tab:3}. We make two observations. First, U-Net3D \cite{b22} fails to achieve good results on RIDER \cite{b28} due to its small size. Second, U-Net3D \cite{b22} degrades in individual dataset performance when trained with multiple datasets, which can be seen by comparing Table \ref{tab:3} and Table \ref{tab:4}.

\begin{table}[htbp]
    \small
    \setlength{\tabcolsep}{4pt} 
    \caption{U-Net3D Segmentation Ablation}
    \begin{center}
    \begin{tabularx}{\linewidth}{|c|c|*{3}{>{\centering\arraybackslash}X|}}
        \hline
        \textbf{Model} & \textbf{Dataset} & \textbf{DSC} & \textbf{IoU} & \textbf{TPF} \\
        \hline
        U-Net3D \cite{b22} & ISPY1 \cite{b8} & 0.52 & 0.40 & 0.42 \\
        U-Net3D \cite{b22} & BreastDM \cite{b45} & 0.62 & 0.52 & 0.65 \\
        U-Net3D \cite{b22} & RIDER \cite{b28} & 0.01 & 0.00 & 1.00 \\
        \hline
        \multicolumn{5}{l}{\parbox{\dimexpr\linewidth-2\tabcolsep}{Ablation study of U-Net3D model performance across BC-MRI-SEG's segmentation datasets. DSC denotes Dice Similarity Coefficient (DSC) score, IoU denotes Intersection over Union, and TPF denotes True Positive Fraction. }} \\
    \end{tabularx}
    \label{tab:3}
    \end{center}
\end{table}

The following analysis is based on stages one and two of our benchmark. Stage one consists of supervised learning and evaluation and is shown in Table \ref{tab:4}. Step two consists of the zero-shot evaluation of trained models, which is shown in Table \ref{tab:5}. Table \ref{tab:5} also presents the scores achieved by random tumor segmentation and classification as a reference.

\begin{table*}[htbp]
    \centering
    \small
    \setlength{\tabcolsep}{4pt} 
    \caption{Supervised Model Performance}
    \begin{tabularx}{\linewidth}{|c|c|*{4}{>{\centering\arraybackslash}X|}c|}
        \hline
        \textbf{MODEL} & \textbf{Combined DSC} & \textbf{ISPY1 DSC} & \textbf{BreastDM DSC} & \textbf{Learnable Param.} & \textbf{Total Param.} \\
        \hline
        U-Net2D \cite{b33} & 0.45 & 0.39 & 0.49 & 1.6M & \textbf{1.6M} \\
        U-Net3D \cite{b22} & 0.46 & 0.35 & 0.54 & 4.8M & 4.8M \\
        U-Net2.1D & 0.49 & 0.34 & 0.60 & 1.6M & 1.6M \\
        SwinUNETR \cite{b17} & 0.70 & 0.62 & 0.75 & 62M & 62M \\
        Med-SA \cite{b42} & 0.72 & \textbf{0.70} & 0.73 & \textbf{1.4M} & 104M \\
        SegResNet \cite{b30} & \textbf{0.75} & 0.68 & \textbf{0.80} & 4.7M & 4.7M \\
        \hline
        \multicolumn{6}{l}{\parbox{\dimexpr\linewidth-2\tabcolsep}{Performance and parameter details of various models when trained on the ISPY1 and BreastDM datasets. DSC denotes Dice Similarity Coefficient (DSC) score, and Param. denotes Parameters. In the BC-MRI-SEG benchmark models are trained and evaluated on the aggregate of the ISPY1 and BreastDM datasets. ISPY1 DSC and BreastDM DSC are the DSC achieved on each individual dataset. Combined DSC is the score achieved when the datasets are aggregated.}} \\
    \end{tabularx}
    \label{tab:4}
\end{table*}

In Table \ref{tab:5}, we see that U-Net3D's \cite{b22} zero-shot performance on RIDER \cite{b28} outperforms its expert counterpart in Table \ref{tab:3}, which shows that the training in stage one allowed for the model to generalize on MRI data. We note that the three U-Net approaches achieve comparable results in both stages of our benchmark. This result is surprising because 3D models are considered superior in the medical imaging literature \cite{b40}. 

\begin{table}[htbp]
    \caption{Zero-shot Model Performance}
    \begin{center}
        \small 
        \setlength{\tabcolsep}{2pt} 
        \begin{tabularx}{\linewidth}{|X|c|c|c|}
            \hline
            \textbf{MODEL} & \textbf{Avg. Score} & \textbf{RIDER DSC} & \textbf{DUKE F1} \\
            \hline
            Random & 0.6 & 0.00 & 0.12 \\
            Med-SA \cite{b42} & 0.17 & 0.17 & 0.16 \\
            U-Net3D \cite{b22} & 0.20 & 0.13 & 0.27 \\
            U-Net2D \cite{b33} & 0.23 & 0.17 & 0.30 \\
            U-Net2.1D & 0.24 & 0.16 & 0.32 \\
            SwinUNETR \cite{b17} & 0.28 & 0.22 & \textbf{0.33} \\
            SegResNet \cite{b30} & \textbf{0.31} & \textbf{0.32} & 0.30 \\
            \hline
            \multicolumn{4}{l}{\parbox{\dimexpr\linewidth-2\tabcolsep}{Zero-shot performance of various models when trained on the ISPY1 and BreastDM datasets and evaluated on the RIDER and DUKE datasets. DSC denotes Dice Similarity Coefficient (DSC) score, Avg. denotes Average, and F1 denotes F1 score. RIDER DSC and DUKE F1 are scores achieved on each individual dataset. Avg. Score is the average of the RIDER DSC and DUKE F1.}} 
        \end{tabularx}
    \end{center}
    \label{tab:5}
\end{table}

Our approach, U-Net2.1D, is three times lower in parameters than U-Net3D \cite{b22}, and outperforms it both in supervised and zero-shot performance. This suggests that the two neighboring two images in an MRI sequence could be sufficient for forming a segmentation mask of a tumor. Furthermore, we hypothesize this would extend to other architectures as well.

In Table \ref{tab:4}, the SOTA models perform comparably. We note that SegResNet \cite{b30} contains significantly fewer total parameters than the other models. However, in Table \ref{tab:5}, we observe that while Med-SA \cite{b42} is competitive in supervised evaluation, adapter-finetuning fails to generalize in a zero-shot setting. Overall, SegResNet \cite{b30} achieves the best performance.

\begin{figure}[htbp]
    \centering
    \includegraphics[width=0.4\textwidth]{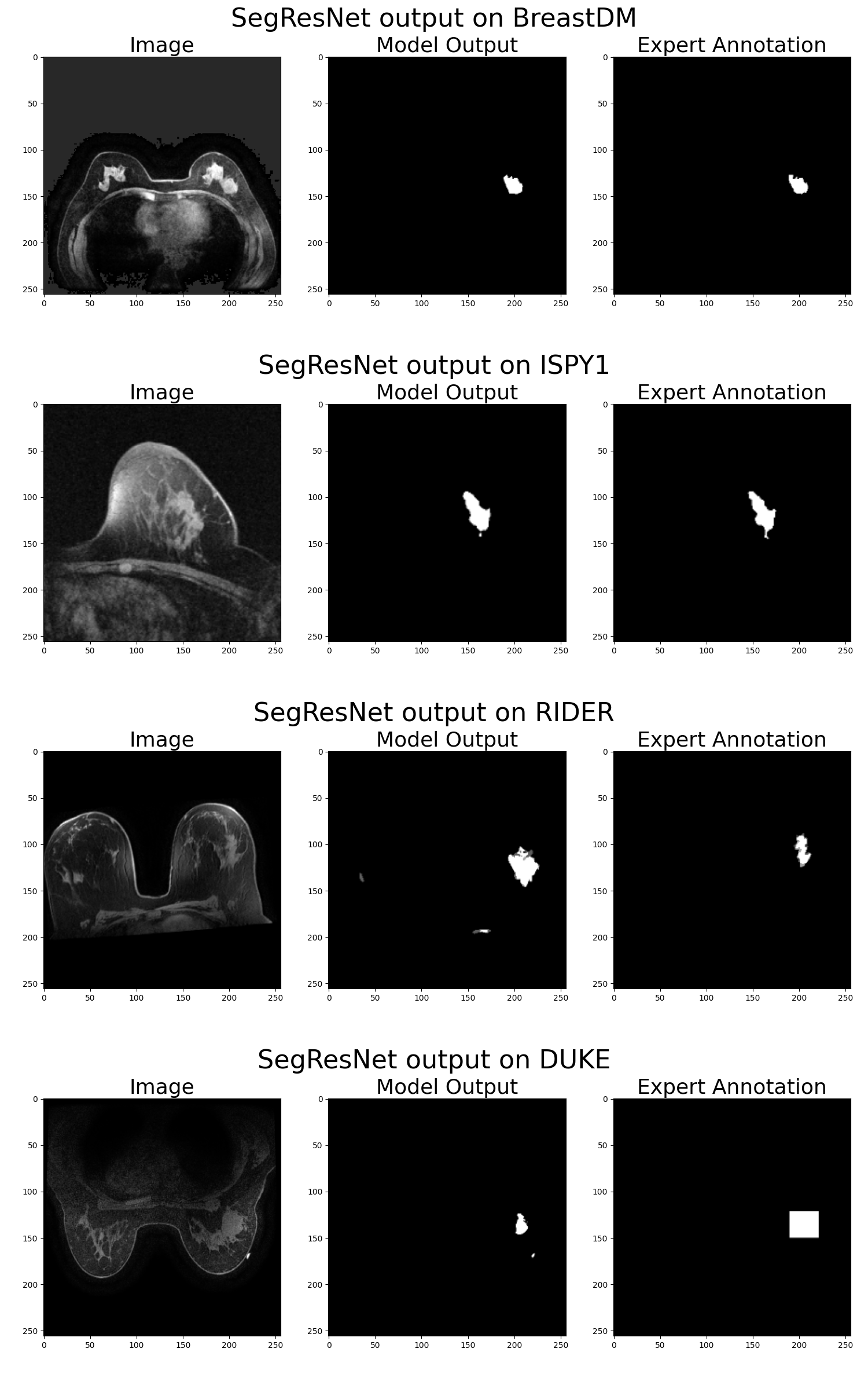}
    \caption{SegResNet Segmentation Mask Outputs.}
    \label{fig}
\end{figure}

In conclusion, our experimental results suggest that adapter-based tuning yields poor zero-shot performance, two neighboring images are sufficient for tumor segmentation, and an asymmetrically large encoder architecture outperforms a traditionally balanced encoder-decoder architecture.



\section{Conclusion and Future Work}
We present BC-MRI-SEG, a benchmark for breast cancer tumor segmentation using publicly available MRI datasets, aiming to inspire the development of accurate and generalizable models for clinical diagnosis and treatment

The datasets, suitable for training larger medical models beyond our benchmark \cite{b44}, offer opportunities such as utilizing pretext tasks \cite{b1}. Additionally, there is potential for fine-grained labeling, including self-supervised learning for creating segmentation masks for DUKE \cite{b34} or enhancing masks in ISPY2 \cite{b25}. Further, extending our U-Net2.1D approach to more complex models and diverse research domains leveraging depth is an avenue for exploration.



\begin{thebibliography}{00}
\bibitem{b1} 
Albelwi, Saleh. "Survey on self-supervised learning: auxiliary pretext tasks and contrastive learning methods in imaging." Entropy 24.4 (2022): 551.
\bibitem{b46}
Antonelli, Michela, et al. "The medical segmentation decathlon." Nature communications 13.1 (2022): 4128.
\bibitem{b2} 
Bertels, Jeroen, et al. "Optimizing the dice score and jaccard index for medical image segmentation: Theory and practice." Medical Image Computing and Computer Assisted Intervention–MICCAI 2019: 22nd International Conference, Shenzhen, China, October 13–17, 2019, Proceedings, Part II 22. Springer International Publishing, 2019.
\bibitem{b3} 
B. H. Menze, A. Jakab, S. Bauer, J. Kalpathy-Cramer, K. Farahani, J. Kirby, et al. "The Multimodal Brain Tumor Image Segmentation Benchmark (BRATS)", IEEE Transactions on Medical Imaging 34(10), 1993-2024 (2015) DOI: 10.1109/TMI.2014.2377694
\bibitem{b4} 
Bloch, B. Nicolas, Jain, Ashali, Jaffe, C. Carl. (2015). BREAST-DIAGNOSIS [Data set]. The Cancer Imaging Archive. http://doi.org/10.7937/K9/TCIA.2015.SDNRQXXR
\bibitem{b5} 
Bouchebbah, Fatah, and Hachem Slimani. "3D automatic levels propagation approach to breast MRI tumor segmentation." Expert Systems with Applications 165 (2021): 113965.
\bibitem{b6} 
C. Gnonnou and N. Smaoui, "Segmentation and 3D reconstruction of MRI images for breast cancer detection," International Image Processing, Applications and Systems Conference, Sfax, Tunisia, 2014, pp. 1-6, doi: 10.1109/IPAS.2014.7043316.
\bibitem{b7} 
Chen, Zhuo, et al. "Understanding the impact of label granularity on cnn-based image classification." 2018 IEEE international conference on data mining workshops (ICDMW). IEEE, 2018.
\bibitem{b8} 
Chitalia, R., Pati, S., Bhalerao, M., Thakur, S., Jahani, N., Belenky, J. V., McDonald, E.S., Gibbs, J., Newitt, D., Hylton, N., Kontos, D.,  Bakas, S. (2021). Expert tumor annotations and radiomic features for the ISPY1/ACRIN 6657 trial data collection [Data set]. The Cancer Imaging Archive. https://doi.org/10.7937/TCIA.XC7A-QT20
\bibitem{b9} 
Chitalia, R., Pati, S., Bhalerao, M., Thakur, S. P., Jahani, N., Belenky, V., McDonald, E. S., Gibbs, J., Newitt, D. C., Hylton, N. M., Kontos, D.,  Bakas, S. (2022). Expert tumor annotations and radiomics for locally advanced breast cancer in DCE-MRI for ACRIN 6657/I-SPY1. In Scientific Data (Vol. 9, Issue 1). Springer Science and Business Media LLC. https://doi.org/10.1038/s41597-022-01555-4
\bibitem{b10} 
Chlap, P., Min, H., Vandenberg, N., Dowling, J., Holloway, L. and Haworth, A. (2021), A review of medical image data augmentation techniques for deep learning applications. J Med Imaging Radiat Oncol, 65: 545-563. https://doi.org/10.1111/1754-9485.13261
\bibitem{b11} 
Clark, K., Vendt, B., Smith, K., Freymann, J., Kirby, J., Koppel, P., Moore, S., Phillips, S., Maffitt, D., Pringle, M., Tarbox, L.,  Prior, F. (2013). The Cancer Imaging Archive (TCIA): Maintaining and Operating a Public Information Repository. In Journal of Digital Imaging (Vol. 26, Issue 6, pp. 1045–1057). Springer Science and Business Media LLC. https://doi.org/10.1007/s10278-013-9622-7 PMCID: PMC3824915
\bibitem{b12} 
Clark, K., Vendt, B., Smith, K., Freymann, J., Kirby, J., Koppel, P., Moore, S., Phillips, S., Maffitt, D., Pringle, M., Tarbox, L.,  Prior, F. (2013). The Cancer Imaging Archive (TCIA): Maintaining and Operating a Public Information Repository. Journal of Digital Imaging, 26(6), 1045–1057. https://doi.org/10.1007/s10278-013-9622-
\bibitem{b13} 
Clunie, D., Hickman, H., Ver Hoef, W., Hastak, S., Evans, J., Neville, J.,  Wagner, U. (2020). Observations from the Data Integration and Imaging Informatics (DI-Cubed) Project. MDPI AG. https://doi.org/10.20944/preprints202008.0474.v1
\bibitem{b14}
D. Yin and R. -W. Lu, "A Method of Breast Tumour MRI Segmentation and 3D Reconstruction," 2015 7th International Conference on Information Technology in Medicine and Education (ITME), Huangshan, China, 2015, pp. 23-26, doi: 10.1109/ITME.2015.117.
\bibitem{b15}
F. Pourpanah et al., "A Review of Generalized Zero-Shot Learning Methods," in IEEE Transactions on Pattern Analysis and Machine Intelligence, vol. 45, no. 4, pp. 4051-4070, 1 April 2023, doi: 10.1109/TPAMI.2022.3191696.
\bibitem{b16}
Haq, Imran Ul, et al. "BTS-GAN: computer-aided segmentation system for breast tumor using MRI and conditional adversarial networks." Engineering Science and Technology, an International Journal 36 (2022): 101154.
\bibitem{b46}
Haralick RM, Stanley RS, Xinhua Z. Image analysis using mathematicalmorphology. IEEE TPAMI 1987;4:532-550.
\bibitem{b17}
Hatamizadeh, A., Nath, V., Tang, Y., Yang, D., Roth, H.R., Xu, D. (2022). Swin UNETR: Swin Transformers for Semantic Segmentation of Brain Tumors in MRI Images. In: Crimi, A., Bakas, S. (eds) Brainlesion: Glioma, Multiple Sclerosis, Stroke and Traumatic Brain Injuries. BrainLes 2021. Lecture Notes in Computer Science, vol 12962. Springer, Cham. https://doi.org/10.1007/978-3-031-08999-2\_22
\bibitem{b18}
Hickman H., Ver Hoef W., Hastak S., Neville J., Clunie D., Wagner U., Helton E. (2019). SDTM datasets of clinical data and measurements for selected cancer collections to TCIA [Dataset]. The Cancer Imaging Archive. doi: 10.7937/TCIA.2019.zfv154m9
\bibitem{b19}
Hicks SA, Strümke I, Thambawita V, Hammou M, Riegler MA, Halvorsen P, Parasa S. On evaluation metrics for medical applications of artificial intelligence. Sci Rep. 2022 Apr 8;12(1):5979. doi: 10.1038/s41598-022-09954-8. PMID: 35395867; PMCID: PMC8993826.
\bibitem{b20}
Jan Witowski et al. ,Improving breast cancer diagnostics with deep learning for MRI.Sci. Transl. Med.14,eabo4802(2022).DOI:10.1126/scitranslmed.abo4802
\bibitem{b21}
J. Qiu et al., "Large AI Models in Health Informatics: Applications, Challenges, and the Future," in IEEE Journal of Biomedical and Health Informatics, vol. 27, no. 12, pp. 6074-6087, Dec. 2023, doi: 10.1109/JBHI.2023.3316750.
\bibitem{b22}
Kerfoot, E., Clough, J., Oksuz, I., Lee, J., King, A.P., Schnabel, J.A. (2019). Left-Ventricle Quantification Using Residual U-Net. In: Pop, M., et al. Statistical Atlases and Computational Models of the Heart. Atrial Segmentation and LV Quantification Challenges. STACOM 2018. Lecture Notes in Computer Science(), vol 11395. Springer, Cham. https://doi.org/10.1007/978-3-030-12029-0\_40
\bibitem{b23}
Kirillov, A., Mintun, E., Ravi, N., Mao, H., Rolland, C., Gustafson, L., ...  Girshick, R. (2023). Segment anything. arXiv preprint arXiv:2304.02643.
\bibitem{b24}
Liu, Ze, et al. "Swin transformer: Hierarchical vision transformer using shifted windows." Proceedings of the IEEE/CVF international conference on computer vision. 2021.
\bibitem{b25}
Li, W., Newitt, D. C., Gibbs, J., Wilmes, L. J., Jones, E. F., Arasu, V. A., Strand, F., Onishi, N., Nguyen, A. A.-T., Kornak, J., Joe, B. N., Price, E. R., Ojeda-Fournier, H., Eghtedari, M., Zamora, K. W., Woodard, S. A., Umphrey, H., Bernreuter, W., Nelson, M., … Hylton, N. M. (2020). Predicting breast cancer response to neoadjuvant treatment using multi-feature MRI: results from the I-SPY 2 TRIAL. In npj Breast Cancer (Vol. 6, Issue 1). Springer Science and Business Media LLC. https://doi.org/10.1038/s41523-020-00203-7
\bibitem{b26}
Li X, Abramson RG, Arlinghaus LR, Kang H, Chakravarthy AB, Abramson VG, Farley J, Mayer IA, Kelley MC, Meszoely IM, Means-Powell J, Grau AM, Sanders M, Yankeelov TE.  Multiparametric magnetic resonance imaging for predicting pathological response after the first cycle of neoadjuvant chemotherapy in breast cancer. Investigative Radiology, 2015 Apr;50(4):195-204. PMCID: PMC4471951
\bibitem{b27}
Mendes, J.; Domingues, J.; Aidos, H.; Garcia, N.; Matela, N. AI in Breast Cancer Imaging: A Survey of Different Applications. J. Imaging 2022, 8, 228. https://doi.org/10.3390/jimaging8090228
\bibitem{b28}
Meyer, C. R., Chenevert, T. L., Galbán, C. J., Johnson, T. D., Hamstra, D. A., Rehemtulla, A.,  Ross, B. D. (2015). RIDER Breast MRI [Data set]. The Cancer Imaging Archive. https://doi.org/10.7937/K9/TCIA.2015.H1SXNUXL
\bibitem{b29}
Morris, E., Burnside, E., Whitman, G., Zuley, M., Bonaccio, E., Ganott, M., Sutton, E., Net, J., Brandt, K., Li, H., Drukker, K., Perou, C.,  Giger, M. L. (2014). Using Computer-extracted Image Phenotypes from Tumors on Breast MRI to Predict Stage [Data set]. The Cancer Imaging Archive. https://doi.org/10.7937/K9/TCIA.2014.8SIPIY6G
\bibitem{b30}
Myronenko, Andriy. "3D MRI brain tumor segmentation using autoencoder regularization." Brainlesion: Glioma, Multiple Sclerosis, Stroke and Traumatic Brain Injuries: 4th International Workshop, BrainLes 2018, Held in Conjunction with MICCAI 2018, Granada, Spain, September 16, 2018, Revised Selected Papers, Part II 4. Springer International Publishing, 2019.
\bibitem{b31}
Newitt, D.,  Hylton, N. (2016). Single site breast DCE-MRI data and segmentations from patients undergoing neoadjuvant chemotherapy (Version 3) [Data set]. The Cancer Imaging Archive. https://doi.org/10.7937/K9/TCIA.2016.QHsyhJKy
\bibitem{b32}
Park, G.E., Kim, S.H., Nam, Y., Kang, J., Park, M. and Kang, B.J. (2024), 3D Breast Cancer Segmentation in DCE-MRI Using Deep Learning With Weak Annotation. J Magn Reson Imaging. https://doi.org/10.1002/jmri.28960
\bibitem{b33}
Ronneberger, Olaf, Philipp Fischer, and Thomas Brox. "U-net: Convolutional networks for biomedical image segmentation." Medical Image Computing and Computer-Assisted Intervention–MICCAI 2015: 18th International Conference, Munich, Germany, October 5-9, 2015, Proceedings, Part III 18. Springer International Publishing, 2015.
\bibitem{b34}
Saha, A., Harowicz, M. R., Grimm, L. J., Kim, C. E., Ghate, S. V., Walsh, R.,  Mazurowski, M. A. (2018). A machine learning approach to radiogenomics of breast cancer: a study of 922 subjects and 529 DCE-MRI features. British journal of cancer, 119(4), 508-516. DOI: https://doi.org/10.1038/s41416-018-0185-8 ,  PMC6134102
\bibitem{b35}
Saha, A., Harowicz, M. R., Grimm, L. J., Weng, J., Cain, E. H., Kim, C. E., Ghate, S. V., Walsh, R.,  Mazurowski, M. A. (2021). Dynamic contrast-enhanced magnetic resonance images of breast cancer patients with tumor locations [Data set]. The Cancer Imaging Archive. https://doi.org/10.7937/TCIA.e3sv-re93
\bibitem{b36}
S. Bakas, H. Akbari, A. Sotiras, M. Bilello, M. Rozycki, J. Kirby, et al., "Segmentation Labels and Radiomic Features for the Pre-operative Scans of the TCGA-GBM collection", The Cancer Imaging Archive, 2017. DOI: 10.7937/K9/TCIA.2017.KLXWJJ1Q
\bibitem{b37}
S. Bakas, H. Akbari, A. Sotiras, M. Bilello, M. Rozycki, J. Kirby, et al., "Segmentation Labels and Radiomic Features for the Pre-operative Scans of the TCGA-LGG collection", The Cancer Imaging Archive, 2017. DOI: 10.7937/K9/TCIA.2017.GJQ7R0EF
\bibitem{b38}
S. Bakas, H. Akbari, A. Sotiras, M. Bilello, M. Rozycki, J.S. Kirby, et al., "Advancing The Cancer Genome Atlas glioma MRI collections with expert segmentation labels and radiomic features", Nature Scientific Data, 4:170117 (2017) DOI: 10.1038/sdata.2017.117
\bibitem{b39}
S. Bakas, M. Reyes, A. Jakab, S. Bauer, M. Rempfler, A. Crimi, et al., "Identifying the Best Machine Learning Algorithms for Brain Tumor Segmentation, Progression Assessment, and Overall Survival Prediction in the BRATS Challenge", arXiv preprint arXiv:1811.02629 (2018)
\bibitem{b40}
Taleb, Aiham, et al. "3d self-supervised methods for medical imaging." Advances in neural information processing systems 33 (2020): 18158-18172.
\bibitem{b41}
Wang, Risheng, et al. "Medical image segmentation using deep learning: A survey." IET Image Processing 16.5 (2022): 1243-1267.
\bibitem{b42}
Wu, Junde, et al. "Medical sam adapter: Adapting segment anything model for medical image segmentation." arXiv preprint arXiv:2304.12620 (2023).
\bibitem{b43}
Yin, X. X., Sun, L., Fu, Y., Lu, R.,  Zhang, Y. (2022). U-Net-Based medical image segmentation. Journal of Healthcare Engineering, 2022.
\bibitem{b44}
Zekai Chen, Devansh Agarwal, Kshitij Aggarwal, Wiem Safta, Mariann Micsinai Balan, Kevin Brown; Proceedings of the IEEE/CVF Winter Conference on Applications of Computer Vision (WACV), 2023, pp. 1970-1980
\bibitem{b45}
Zhao X, Liao Y, Xie J, He X, Zhang S, Wang G, Fang J, Lu H, Yu J. BreastDM: A DCE-MRI dataset for breast tumor image segmentation and classification. Comput Biol Med. 2023 Sep;164:107255. doi: 10.1016/j.compbiomed.2023.107255. Epub 2023 Jul 10. PMID: 37499296.
\end{thebibliography}
\end{document}